\documentclass[letterpaper, 10 pt, conference]{ieeeconf}  % Comment this line out % if you need a4paper
\pdfoutput=1

\IEEEoverridecommandlockouts  % This command is only needed if you want to use the \thanks command

\usepackage{graphicx}
\usepackage{tikz}
\usepackage{subcaption}
\usepackage{pgfplots}
\pgfplotsset{width=.9\linewidth,compat=1.13,
                    every axis/.append style={
                                label style={font=\small},
                                tick label style={font=\small},
                                enlarge x limits={abs=0.25}},
                                /pgfplots/ybar legend/.style={
                                /pgfplots/legend image code/.code={%
                                \draw[##1,/tikz/.cd,yshift=-0.25em]
                                (0cm,0cm) rectangle (3pt,0.8em);},
},
}

\usepackage[ruled,norelsize]{algorithm2e}
\title{\LARGE \bf
Analyzing Adversarial Attacks Against Deep Learning for Intrusion Detection in IoT Networks}

\author{ \parbox{5 in}{\centering Olakunle Ibitoye, Omair Shafiq and Ashraf Matrawy 
         \thanks{}\\
         School of Information Technology\\
         Carleton University, Ottawa, Canada\\
         \tt\small Email: \{Kunle.Ibitoye, Omair.Shafiq, Ashraf.Matrawy\}@carleton.ca
}
}

\begin{document}

\maketitle
\thispagestyle{empty}
\pagestyle{empty}

\begin{abstract}

    % Deep learning techniques such as Feed-forward Neural Networks (FNN) have been widely used to detect and mitigate security threats against IoT networks. However, deep learning models have been shown to be vulnerable to adversarial attacks. In this paper, we consider a variant of the FNN known as the Self-normalizing neural network (SNN) and compare its performance with the FNN for classifying intrusion attacks in an IoT network. In our experimental results, the FNN outperforms the SNN based on multiple performance metrics such as accuracy, precision and recall. However, when tested for adversarial robustness, the SNN demonstrates greater resilience against several forms of adversarial attacks. This study presents a promising future in the quest for safer and more secure deep learning in IoT networks.
    
    Adversarial attacks have been widely studied in the field of computer vision but their impact on network security applications remains an area of open research. As IoT, 5G and AI continue to converge to realize the promise of the fourth industrial revolution (Industry 4.0), security incidents and events on IoT networks have increased. Deep learning techniques are being applied to detect and mitigate many of such security threats against IoT networks. Feed-forward Neural Networks (FNN) have been widely used for classifying intrusion attacks in IoT networks. In this paper, we consider a variant of the FNN known as the Self-normalizing Neural Network (SNN) and compare its performance with the FNN for classifying intrusion attacks in an IoT network. Our analysis is performed using the BoT-IoT dataset from the Cyber Range Lab of the center of UNSW Canberra Cyber. In our experimental results, the FNN outperforms the SNN for intrusion detection in IoT networks based on multiple performance metrics such as accuracy, precision, and recall as well as multi-classification metrics such as Cohen Cappa’s score. However, when tested for adversarial robustness, the SNN demonstrates better resilience against the adversarial samples from the IoT dataset, presenting a promising future in the quest for safer and more secure deep learning in IoT networks. 


\end{abstract}

\section{Introduction}
As the Internet of Things (IoT) emerges and expands over the next several years, the security risks in IoT will increase. There will be bigger rewards for successful IoT breaches and hence greater incentive and motivation for attackers to find new and novel ways to compromise IoT systems. Traditional methods and techniques for protecting against cyber threats in the traditional internet will prove inadequate in protecting against the unique security vulnerabilities that would be expected in the internet of things \cite{roman2011securing}. Hence security researchers and professionals would need to evaluate existing processes and improve upon them to create more efficient security solutions to address the security vulnerabilities in the emerging Internet of Things.

Managing security challenges in any network involves three broad strategies namely prevention, detection and mitigation. Successful security solutions for IoT networks will need to adopt all three measures. For the scope of this paper, we focus on Intrusion Detection Systems (IDS) and consider deep learning based IDS for detecting and classifying network traffic within an IoT enviroment. 

% RFC 7452 \cite{tschofenig2015architectural} differentiates between the various types of end devices  in a computer network. End devices such as personal computers, servers, tablets and smart phones which are designed for general purpose computing are called computing devices. For the scope of this study, we refer to a network that was designed for and consists mostly of this type of end devices as a "Traditional network". End devices such as light bulbs, smart speakers, smart door bells which are presumable designed for limited or specific purposes are called "IoT devices". In our study, we refer to a network consisting mostly of such end devices as an "IoT network". 

% Deep learning based IDS are more efficient than signature based IDS in detecting zero day attacks. 
Deep learning based IDS have an advantage over conventional anomaly based IDS because they help overcome the challenge of proper feature selections \cite{javaid2016deep}. However, two major challenges of deep learning in security applications are the lack of transparency of the deep learning models \cite{guo2018lemna}, and the vulnerability of the deep learning models to adversarial attacks \cite{Szegedy2014IntriguingPO}. For the scope of this study, we focus on adversarial vulnerability of the deep learning models.

An adversarial attack occurs when an adversarial example is fed as an input to a machine learning model. An adversarial example is an instance of the input in which some feature has been intentionally perturbed with the intention of confusing a machine learning model to produce a wrong prediction. Szegedy et al. \cite{Szegedy2014IntriguingPO} demonstrated how a deep learning model for image recognition could be confused into making wrong predictions by introducing a tiny perturbation to the image. Other researchers \cite{papernot2016limitations} \cite{wang2018deep} have also proved that adversarial attacks are equally effective against deep learning models in network security applications such as malware detection and intrusion detection systems.

A Self-normalizing Neural Network (SNN) \cite{klambauer2017self} is a type of deep learning model that maintains the stability of the network during the gradient descent process. Since there is currently no research, to the best of our knowledge, that has been carried out to analyze adversarial attacks against Self-normalizing Neural Networks for intrusion detection in IoT networks, we propose to carry out our study to close out this gap.

\textbf{Our Contributions} in this paper are as follows: For our \textbf{ first contribution}, this is to the best of our knowledge, the first study to demonstrate the effects of adversarial samples on a deep learning based Intrusion Detection System (IDS) within the context of an IoT network. \textbf{For our second contribution} we provide a comparison between the performance of an IDS implemented with two different deep learning models - a Self-normalizing Neural Network (SNN) and a typical Feed-forward Neural Network (FNN) within the context of an IoT network. In our \textbf{third contribution}, we demonstrate that while the IDS implemented with FNN performs better than the SNN based IDS with regards to performance metrics such as accuracy, precision and recall, the SNN based IDS is however more robust to adversarial samples. In our \textbf{fourth and final contribution}, we analyze the effects of feature normalization on the adversarial robustness of deep learning based IDS in IoT. This is the first study to the best of our knowledge to demonstrate that normalization of input features in a deep learning based IDS adversely impacts the ability of the deep learning model to resist adversarial attacks.

\section{Related Work}

While previous research \cite{wang2018deep} have utilized deep learning techniques for intrusion detection in traditional networks, in this study, we extend this research area by specifically applying deep learning for intrusion detection in the context of IoT. We then demonstrate that deep learning models used for intrusion detection in IoT can be confused with adversarial samples.

Moustafa et al \cite{Koroniotis2018TowardsTD} in the original paper that described the IoT dataset that we used for our experiments implemented LSTM, SVM and RNN machine learning techniques to analyze the IoT dataset but they did not evaluate the adversarial robustness of their machine learning models in their study. Additionally, their study only carried out binary classification on the dataset and the prediction output of the machine learning models was classified as either attack or normal traffic. We note that the usefulness of such studies prevails for network forensic analysis use cases where it is essential to classify the output into the various categories of attacks. 

Hodo et al. \cite{hodo2016threat} analyzed the threat of intrusion detection against IoT using a very limited dataset sample of  2313 training samples, 496 validation samples and 496 test samples. The dataset also contains only DDoS/DoS traffic and normal traffic. In a realistic IoT environment, we expect a larger network traffic dataset with hundreds of thousands or millions of records with a more heterogeneous attack profile on the network. We used a dataset containing over 3.6 million records and a more heterogeneous attack profile consisting of 5 target labels.

Zheng \cite{wang2018deep} implemented several adversarial attack algorithms against a deep learning based intrusion detection system in a traditional network using multi-layer perceptron Feed-forward Neural Network and compared the results from the various adversarial attacks. The author demonstrated that the deep learning based IDS classifier using the FNN was adversely impacted  by the adversarial samples. However, the NSL-KDD dataset that was used was generated over a decade ago and may not represent the type of network attack traffic that would be expected in today's IoT networks. Warzynski et al. \cite{warzynski2018intrusion} also evaluated the NSL-KDD dataset by training a FNN to classify the network packets, and then tested the resilience of their model to adversarial examples. The dataset used in their experiment may not represent a typical IoT network traffic.

Based on our literature review and study of related work, we discovered that no researcher has evaluated the resilience of Self-normalizing Neural Networks (SNN) to adversarial examples for deep learning based IDS in IoT networks. Hence our study is novel and offers a useful contribution in understanding the security of machine learning and artificial intelligence in IoT.

\section{Problem Definition and Proposed Study}

Zheng \cite{wang2018deep} demonstrated that a deep learning based IDS that could correctly identify DoS attacks with an accuracy of 93\% could have its performance degraded to as low as 24\% with adversarial samples. The deep learning model used in the study was a Feed-forward Neural Network (FNN).

FNN with deep architectures have been known to experience gradient decay, resulting in poor performance. Klambauer et al \cite{klambauer2017self} proposed the Self-normalizing Neural Networks (SNN) which is a variant of the FNN that uses a Scaled Exponential Linear Unit (SELU) activation function.

% alongside a custom method for initializing the weights in the neural network. The scaled exponential linear unit is shown as:

% \[selu(x) = \left\{
%                 \begin{array}{ll}
%                   x & if \, x >0\\ 
%                   \alpha^x - \alpha & if \; x <= 0\\
%                 \end{array}.
%               \right.\\
%               \]

% With the SeLU activation function, the mean of the activation output is kept at zero and the variance is kept at one. This allows of deeper neural network architectures to be trained without suffering significant gradient decay. 

The performance of SNN in the context of intrusion detection in IoT in currently not known as well as their resilience to adversarial samples. Since no study, to the best of our knowledge, has tested the resilience of SNN for intrusion detection in IoT, we propose to carry out our study to close out this gap. 

\section{Experimental Approach}
In carrying out this study, we implement deep learning based intrusion detection systems for an IoT dataset. We then test the resilience of the deep learning IDSes to adversarial samples. To demonstrate this, we create our own adversarial samples from the IoT dataset which we used in training the deep learning models. The methods used in crafting the adversarial samples for this study are the Fast Gradient Sign Method (FGSM) \cite{Goodfellow2015ExplainingAH}, Basic Iteration Method (BIM)\cite{kurakin2016adversarial}, and the Projected Gradient Descent\cite{madry2018towards}. Our study considers the following \textbf{assumptions}. The attacks are evasion attacks which are launched during the prediction phase of the deep learning model. Also, a complete knowledge of the deep learning model is assumed, hence they are white box attack. In this study, we do not target any specific prediction outcome, rather we seek to confuse the deep learning classifier to make a mistake and produce a misclassification. Hence a reliability attack. Our expected outcome is to degrade the performance of deep learning classifier, as measured by various performance metrics.

\begin{figure}[h]
	\includegraphics[width=\linewidth,keepaspectratio=true]{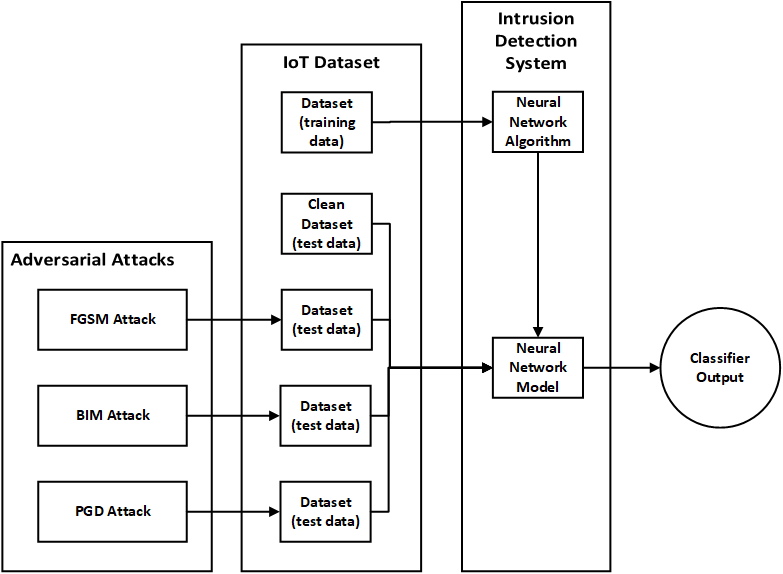}
	\caption{Solution Overview Architecture}
	\label{fig:adv_attack}
	\centering
\end{figure}

 \subsection{Development platform and tools}
We develop our deep learning code in python language with jupyter notebook hosted in Google Colaboratory. Colaboratory is an interactive environment provided by Google for writing and executing code in python and other languages \cite{exp}. Colaboratory offers advanced GPU features, is hosted in the cloud, and comes with several pre-installed deep learning frameworks and libraries that accelerate the task of building machine learning models. 

\subsection{Dataset}

For our dataset, we use the BoT-IoT dataset \cite{Koroniotis2018TowardsTD} provided from the Cyber Range Lab of The center of UNSW Canberra Cyber. This dataset provides a realistic representation of an IoT network since it was created in a dedicated IoT environment, and contains adequate number of records with heterogeneous network profiles.

The BoT IoT  dataset consists over \textbf{72 million records} of network activity in a simulated IoT environment. A scaled-down version of the dataset comprising of approximately \textbf{3.6 million records} is also available and was used for our study. A selection of the 10 best features have been provided in the original dataset and were also used for this study.  \cite{Koroniotis2018TowardsTD}.

The training and test dataset have 5 output classes each which reflect the normal traffic and the 4 types of attacks which were carried out against the IoT network.

\begin{table}[ht]
\caption{BoT-IoT Dataset Features}
\label{table_attack_ml}
\begin{center}
\begin{tabular}{|c|c|}
\hline
Feature &  Description \\
\hline
\hline
pkSeqID & Row Identifier\\
\hline
Stime & Record start time\\
\hline
Seq  & Argus sequence number\\
\hline
Mean & Average duration of aggregated records\\
\hline
Stddev  & Standard deviation of aggregated records\\
\hline
Min & Minimum duration of aggregated records\\
\hline
Max  & Maximum duration of aggregated records\\
\hline
Srate & Source-to-destination packets per second\\
\hline
Drate  & Destination-to-source packets per second\\
\hline
N\_IN\_Conn\_P\_SrcIP & Total Number of packets per source IP\\
\hline
N\_IN\_Conn\_P\_DstIP & Total Number of packets per Destination IP\\
\hline
\end{tabular}
\end{center}
\end{table}

\begin{table}[ht]
\caption{BoT-IoT Dataset Target Classes}
\label{table_attack_ml}
\begin{center}
\begin{tabular}{|c|c|c|}
\hline
Target Label & Training Samples & Test Samples \\
\hline
\hline
DDoS & 1541315 & 385309\\
\hline
DoS & 13201485 & 330112\\
\hline
Reconnaisance & 72919 & 18163\\
\hline
Normal & 370 & 107\\
\hline
Theft & 65 & 14\\
\hline

\end{tabular}
\end{center}
\end{table}

\subsection{Building the FNN and SNN deep learning based IDS}

We implement two IDSes for our IoT dataset. The first IDS is implemented using a Feed Forward Artificial Neural Network (FNN) as shown in fig. 2 while the second IDS is implemented using a Self-normalizing Neural Network (SNN) as shown in fig. 3. In each Neural Network model design, we create 3 hidden layers and 16 neurons for each layer, giving us a total of 48 neurons in the hidden layers. 

The intuition behind SNN is to keep the mean and the variance as close to 0 and 1 respectively throughout each layer of the neural network. 

For the SNN, we use a Scaled Exponential Linear Unit (SeLU) activation function while for the FNN we use a Rectifier Linear Unit(ReLU). The FNN uses a basic dropout layer to prevent overfitting and ensure better stability in the network during the learning phases while the SNN uses an AlphaDropout layer to retain the mean and variance at 0 and 1 respectively. 

For initializing the weights, we select Glorot Uniform initializer \cite{glorot2010understanding} for the FNN while we use a Lecun Uniform Initializer \cite{lecun2012efficient} for the SNN. 

% For both FNN and SNN neural network models, we select the Adaptive Moment Estimation (Adam) optimization algorithm built in to Keras. Compared to other optimization methods, Adam calculates adaptive learning rates during the network learning process for each of the parameters. 

\begin{figure}[h]
	\includegraphics[width=\linewidth,keepaspectratio=true]{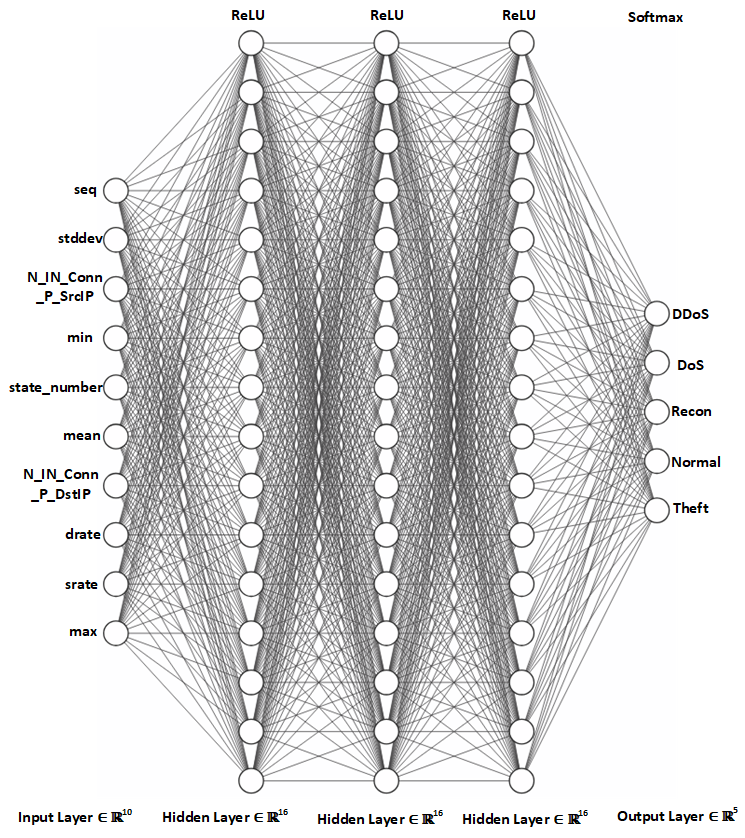}
	\caption{FNN Architecture}
	\label{fig:ML_input}
	\centering
\end{figure}

\begin{figure}
	\includegraphics[width=\linewidth,keepaspectratio=true]{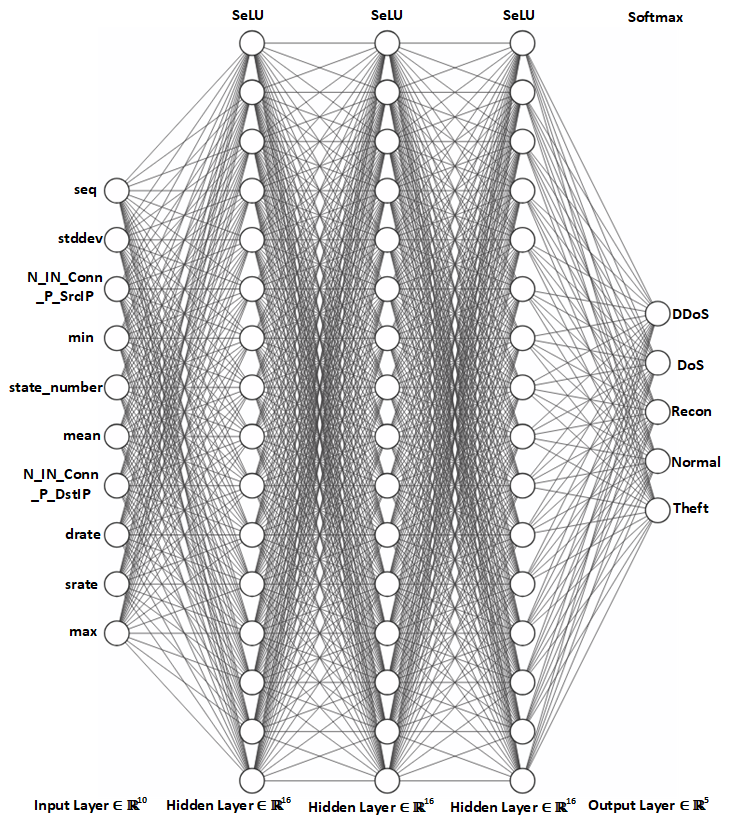}
	\caption{SNN Architecture}
	\label{fig:ML_input}
	\centering
\end{figure}

\subsection{Generating The Adversarial Samples}
We generate our adversarial samples using the Adversarial Robustness Toolbox (ART) \cite{nicolae2018adversarial} framework which is provided by IBM and is made available for public use.

The first method we use in generating the adversarial examples for the IoT dataset is the Fast Gradient Sign Method (FGSM). This method performs a one step gradient update along the direction of the sign of gradient for every input in the dataset. \cite{Goodfellow2015ExplainingAH}. The second method is the Basic Iteration method (BIM) which runs a finer optimization of the FGSM with minimal smaller changes for multiple iterations \cite{kurakin2016adversarial}. In each iteration, the each feature of the input values is clipped to avoid too large a change on each feature. The third method is the Projected Gradient Descent (PGD) which is also a variation of the FGSM attack but omits the random start feature of the FGSM \cite{madry2018towards}. All three methods are model dependent methods\cite{Gong2017AdversarialAC} and rely on the model gradient.

% In  our experiments, we craft the adversarial examples with the intent of misleading the classifier without any specific target labels being specified. The epsilon number represents the maximum perturbation for the adversarial attack and in our experiment, we select an epsilon of 0.3. The algorithm below illustrates the steps taken to build the model, generate the adversarial samples, and test the resilience of the IoT deep learning classifier.

\makeatletter
\newcommand{\removelatexerror}{\let\@latex@error\@gobble}
\makeatother

\begin{figure}
 \removelatexerror
  \begin{algorithm}[H]
   \caption{Adversarial Testing for FNN and SNN}
 \For{Each neural network FNN, SNN}
   {
  initialize {number of hidden layers L, weights w}\;
  Add  input layer, activation layer, dropout layer\\
  
  \For {i in L - 1}
  {
  Add Dense layer\\
\eIf{ANN model}{
Add ReLU activation layer\;
Add Dropout Layer\;}
{ Add SeLU activation Layer\;
Add AlphaDropout Layer\;}

  }
  Add output layer, softmax activation layer, dropout layer\\
    \While{no of epochs not complete}{
compute training and validation loss\;}
  Evaluate model performance\\

          \For{each adversarial attack method}
        {
            Craft adversarial samples x' \\
            Evaluate model performance with adversarial samples x'

        }
  
  {

  }      
}
  \end{algorithm}
\end{figure}

\section{Results \& Evaluation}
Our first result in subsection (A) below illustrates the impact of adversarial samples on a deep learning based IDS implemented using a FNN for the IoT dataset used in this paper. In our second result in subsection (B), we provide a performance comparison between the SNN and the FNN IDSes. In our third result in subsection (C), we compare the adversarial resilience of both the FNN and the SNN IDSes. Our final evaluation in subsection (D) shows the effect of feature normalization on deep learning based IDS using the IoT dataset.

\subsection{Effect of Adversarial Samples on Deep learning based IDS in IoT networks} 
In our first experiment, we demonstrate that the FNN deep learning based IDS was significantly degraded by the adversarial samples generated from the IoT dataset. After training the IDS model, we achieve an initial accuracy of 95.1\%. We then evaluate the performance of the IDS once again using the three adversarial sample datasets that were created in the previous section. The prediction accuracy of the FNN IDS is reduced from 95.1\% to 24\% from the FGSM adversarial samples. We repeat the experiment with the BIM and PGD adversarial samples and achieve accuracies of 18\% and 31\% respectively as shown in Fig 4.

\pgfplotstableread[row sep=\\,col sep=&]{
	K    & Accuracy\\
	1   & 95 \\
	2   & 24\\
	3  & 18 \\
	4 & 31\\
}\mydataSliceAllocated

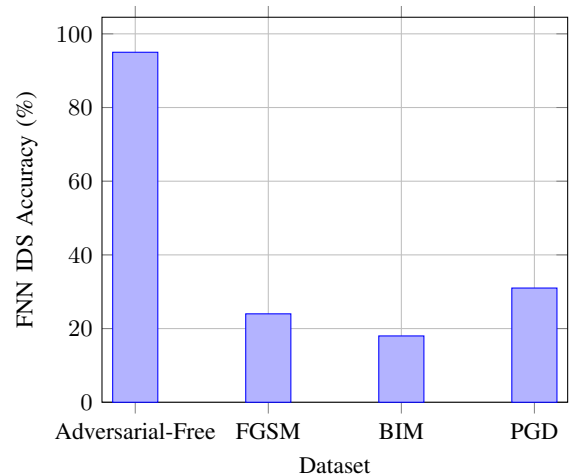
\begin{figure}[ht!]
	\centering
	\begin{tikzpicture}[baseline]
	\begin{axis}[
	ybar,
	ymin=0,
	grid=major,
	ylabel={FNN IDS Accuracy (\%)},
	xlabel={Dataset},
	%yticklabels from table={\mydataSliceAllocated}{1},                
	%xticklabel style={rotate=90,anchor=east},
	xticklabels={Adversarial-Free,FGSM,BIM, PGD},
%	legend style={at={(1,2)},
%		anchor=north,legend columns=-1},
%	legend style={font=\scriptsize,at={(0.5,-0.24)},
%		anchor=north,legend columns=-1},        
	%symbolic x coords={Clean Dataset,FGSM,BIM, PGD},
	xtick=data,
	bar width=6mm,
	%nodes near coords,
	]
	\addplot table[x=K,y=Accuracy]{\mydataSliceAllocated};
	% \addplot table[x=K,y=FCFS]{\mydataSliceAllocated};
	%\legend{DSAF, FCFSFA}
	\end{axis}
	\end{tikzpicture}
	\caption{Effect of Adversarial Samples on FNN IDS}
	\label{fig:ASDLB}
\end{figure}

\subsection{Performance Comparison of FNN and SNN IDS using the adversarial-free IoT dataset}
In our second experiment, we compare both the FNN and SNN IDSes. The SNN IDS underperforms the FNN IDS based on several performance metrics as shown in Fig 5. For classification metrics namely precision, recall and F1-score, the FNN IDS consistently  outperforms the SNN IDS over multiple experiment runs. For additional multiclassification metrics such Copen Cappa Score and MC Coefficient, the FNN IDS also outperforms the SNN IDS as shown in Fig 5.

\pgfplotstableread[row sep=\\,col sep=&]{
	Metric    & FNN & SNN\\
	1   & 95 & 91\\
	2   & 95 & 92\\
	3  & 95 & 91 \\
	4 & 95 & 91\\
	5 & 89 & 84\\
	6 & 89 & 85\\
}\mydataModelAccuracy

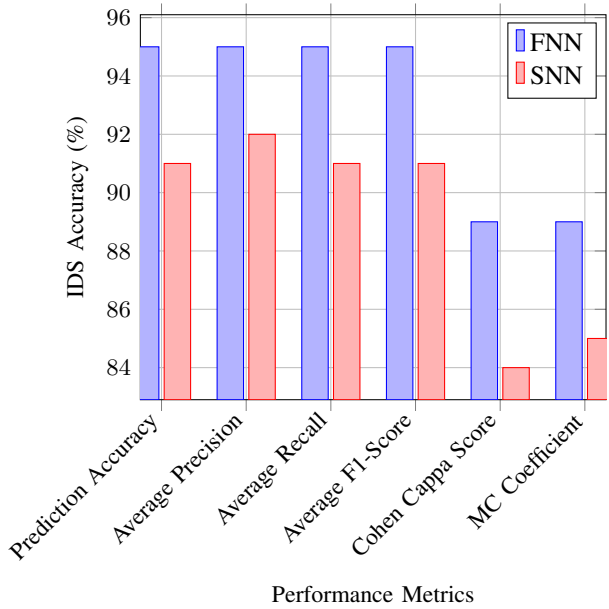
\begin{figure}[ht!]
	\centering
	\begin{tikzpicture}[baseline]
	\begin{axis}[
	ybar,
	grid=major,
	ylabel={IDS Accuracy (\%)},
	xlabel={Performance Metrics},
	%yticklabels from table={\mydataSliceAllocated}{1},                
	xticklabel style={rotate=45,anchor=east},
	xticklabels={Prediction Accuracy,Average Precision,Average Recall, Average F1-Score,Cohen Cappa Score,MC  Coefficient},
	xtick=data,
	%bar width=6mm,
	%nodes near coords,
	]
	\addplot table[x=Metric,y=FNN]{\mydataModelAccuracy};
	\addplot table[x=Metric,y=SNN]{\mydataModelAccuracy};
	\legend{FNN, SNN}
	\end{axis}
	\end{tikzpicture}
	\caption{Comparison of FNN and SNN IDSes}
	\label{fig:Accuracy_Comparison}
\end{figure}

\subsection{Comparison of Adversarial Resilience of FNN IDS and SNN IDS}

Both the FNN IDS and SNN IDS performance on the IoT dataset were degraded by the adversarial samples which we created. We however observe that the SNN IDS is more resilient to the adversarial attacks than the FNN IDS as shown in Fig. 6.

\pgfplotstableread[row sep=\\,col sep=&]{
	Metric    & FNN & SNN\\
	1   & 95 & 91\\
	2   & 24 & 32\\
	3   & 18 & 26\\
	4  & 31 & 42 \\
}\mydataModelPrediction

\begin{figure}[ht!]
	\centering
	\begin{tikzpicture}[baseline]
	\begin{axis}[
	ybar,
	ymax=100,
	grid=major,
	ylabel={IDS Accuracy (\%)},
	xlabel={Performance Metrics},
	%yticklabels from table={\mydataSliceAllocated}{1},   
%    legend style={
%		anchor=east,legend columns=1},
%	legend style={font=\scriptsize,at={(0.5,-0.24)},
%		anchor=north,legend columns=-1},              
	xticklabel style={rotate=45,anchor=east},
	xticklabels={Adversarial-free, FGSM Attack, BIM Attack, PGD Attack},
	xtick=data,
	%bar width=6mm,
	%nodes near coords,
	]
	\addplot table[x=Metric,y=FNN]{\mydataModelPrediction};
	\addplot table[x=Metric,y=SNN]{\mydataModelPrediction};
	\legend{FNN, SNN}
	\end{axis}
	\end{tikzpicture}
	\caption{Adversarial Resilience of FNN and SNN IDS models}
	\label{fig:ModelPrediction}
\end{figure}
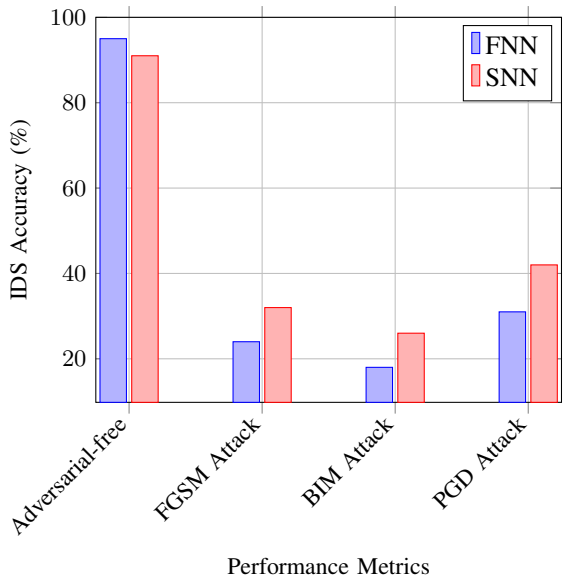

\subsection{Effect of Feature Normalization on a Deep Learning based IDS for IoT}
% Feature normalization is a common pre-processing step in deep learning applications for transforming input features to zero mean and unit variance. In our previous experiments, the input features were normalized during the data-preprocessing phase using the min-max scaler class in scikitlearn python library. 

In our final experiment, we refrain from carrying out feature normalization on the IoT dataset. As shown in Fig. 7 \& 8, both IDSes have a significantly lower prediction accuracy on the adversarial-free dataset when the input features are not normalized. However, their resilience to adversarial samples is improved.

Fig 9 compares the effect of feature normalization on various classification metrics for both FNN and SNN based IDSes using the adversarial-free dataset. The results indicate that both IDSes yield more accurate results on the adversarial-free dataset when the input features are normalized.

\pgfplotstableread[row sep=\\,col sep=&]{
	Metric    & Normalized & Raw Input\\
	1   & 95 & 52\\
	2   & 24 & 51\\
	3  & 18 & 52 \\
	4  & 31 & 52 \\
}\mydataModelPrediction

\begin{figure}[ht!]
	\centering
	\begin{tikzpicture}[baseline]
	\begin{axis}[
	ybar,
	ymax=100,
	grid=major,
	ylabel={FNN IDS Accuracy (\%)},
	xlabel={Dataset},
	%yticklabels from table={\mydataSliceAllocated}{1},   
%    legend style={
%		anchor=east,legend columns=1},
%	legend style={font=\scriptsize,at={(0.5,-0.24)},
%		anchor=north,legend columns=-1},              
	xticklabel style={rotate=45,anchor=east},
	xticklabels={Adversarial-free, FGSM Attack, BIM Attack, PGD Attack},
	xtick=data,
	%bar width=6mm,
	%nodes near coords,
	]
	\addplot table[x=Metric,y=Normalized]{\mydataModelPrediction};
	\addplot table[x=Metric,y=Raw Input]{\mydataModelPrediction};
	\legend{Normalized, Raw input}
	\end{axis}
	\end{tikzpicture}
	\caption{Effect of Feature Normalization on Deep learning based IDS using FNN}
	\label{fig:ModelPrediction}
\end{figure}

\pgfplotstableread[row sep=\\,col sep=&]{
	Metric    & Normalized & Raw Input\\
	1   & 91 & 52\\
	2   & 32 & 51\\
	3  & 26 & 52 \\
	4  & 42 & 52 \\
}\mydataModelPrediction

\begin{figure}[h!]
	\centering
	\begin{tikzpicture}[baseline]
	\begin{axis}[
	ybar,
	ymax=100,
	grid=major,
	ylabel={SNN IDS Accuracy (\%)},
	xlabel={Dataset},
	%yticklabels from table={\mydataSliceAllocated}{1},   
%    legend style={
%		anchor=east,legend columns=1},
%	legend style={font=\scriptsize,at={(0.5,-0.24)},
%		anchor=north,legend columns=-1},              
	xticklabel style={rotate=45,anchor=east},
	xticklabels={Adversarial-free, FGSM Attack, BIM Attack, PGD Attack},
	xtick=data,
	%bar width=6mm,
	%nodes near coords,
	]
	\addplot table[x=Metric,y=Normalized]{\mydataModelPrediction};
	\addplot table[x=Metric,y=Raw Input]{\mydataModelPrediction};
	\legend{Normalized, Raw input}
	\end{axis}
	\end{tikzpicture}
	\caption{Effect of Feature Normalization on Deep learning based IDS using SNN}
	\label{fig:ModelPrediction}
\end{figure}

\begin{figure*}
	\includegraphics[width=.9\linewidth,keepaspectratio=true]{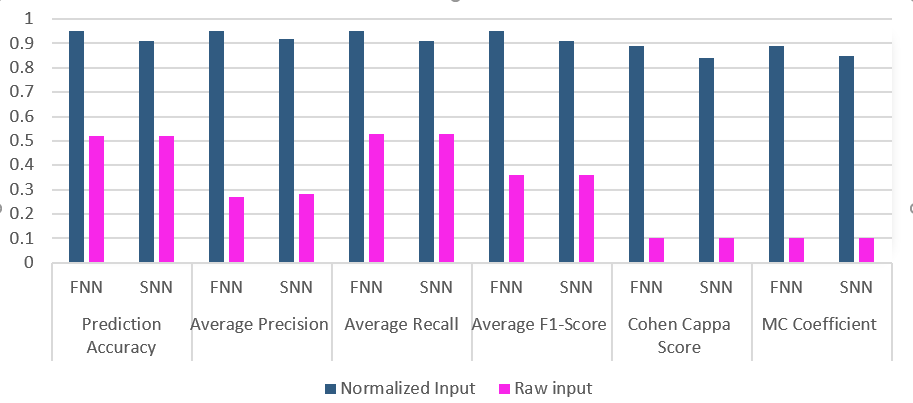}
	\caption{Effect of Feature Normalization in Deep Learning Based IDS in IoT Networks}
	\label{fig:Accuracy_Comparison}
	\centering
\end{figure*}

\section{CONCLUSION}

 We observe from our experiments using an IoT dataset that adversarial samples are a real threat for deep learning based IDS in IoT. 
 
 We created two deep learning based IDSes for an IoT dataset using two types of neural network models - a FNN and a SNN - and observed that both models are impacted differently by the adversarial samples. Our experiments with the IoT dataset show that the self-normalizing feature of the SNN makes it more resilient to gradient based adversarial samples.  

Our results further show that feature normalization of the IoT dataset negatively affects the adversarial resilience of the deep learning based IDSes. When the input features are normalized, both IDSes have better performance metrics, but they are more vulnerable to adversarial samples. 

For future work, we would like to investigate why the self-normalizing properties of the SNN makes the SNN IDS for the IoT dataset more resilient to adversarial samples.

\section*{acknowledgement}
This work was supported by the Natural Sciences and Engineering Research Council of Canada (NSERC) through the NSERC Discovery Grant program.

\bibliographystyle{ieeetr}
\bibliography{main}

% %Where the bibliography will be printed
%   \printbibliography

\end{document}